\pdfoutput=1
\documentclass[aps,prl,floatfix,twocolumn,reprint,superscriptaddress]{revtex4-1}
\usepackage{amsfonts}
\usepackage{mathrsfs}
\usepackage{amsmath}
\usepackage{color}
\usepackage{graphicx}
\usepackage{bm}
\usepackage{amssymb}
\usepackage{xspace}
\usepackage{epstopdf}
\usepackage{dcolumn}
\usepackage{longtable}
\usepackage{multirow}
\usepackage[colorlinks=true, letterpaper=true, pdfstartview=FitV, linkcolor=blue, citecolor=blue, urlcolor=blue]{hyperref}

\makeatletter

\newcommand{\Rmnum}[1]{\expandafter\@slowromancap\romannumeral #1@}
\makeatother

\begin{document}
\title{Circular Dichroism and Radial Hall Effects in Topological Materials}
\author{Ying Liu}
\affiliation{Research Laboratory for Quantum Materials, Singapore University of Technology and Design, Singapore 487372, Singapore}
\author{Shengyuan A. Yang}\email{shengyuan\_yang@sutd.edu.sg}
\affiliation{Research Laboratory for Quantum Materials, Singapore University of Technology and Design, Singapore 487372, Singapore}
\author{Fan Zhang}\email{zhang@utdallas.edu}
\affiliation{The Department of Physics, The University of Texas at Dallas, Richardson, Texas 75080, USA}
\begin{abstract}
Under symmetry breaking, a three-dimensional nodal-line semimetal can turn into
a topological insulator or Weyl semimetal, accompanied by the generation of momentum-space Berry curvature.
We develop a theory that unifies their circular dichroism
and highlights the roles of Berry curvature distribution and light incident direction.
Nontrivially, these phases exhibit distinct dichroic optical absorption and radial Hall effects,
with characteristic scalings with photon energy and electric field.
Our findings offer a new diagnosis tool for examining topological phases of matter.
\end{abstract}
\maketitle

Topological phases of matter have been
attracting significant attention in the past decade~\cite{Kane,Qi,Chiu2016,Volovik}.
A variety of topological insulators (TI) and semimetals have been discovered successively,
which often feature protected band crossing points.
For instance, TI's possess protected Kramers degeneracies on their surfaces~\cite{Kane,Qi,Chiu2016},
whereas Dirac and Weyl semimetals have discrete point singularities in their bulk~\cite{Volovik,Murakami,Wan,Balents,A1,A2,Young,A3,A4,A5,Zhao2013,BJYang,Yang2014}.
Intriguingly, the band-crossing can also form a symmetry-protected 1D loop~\cite{Yang2014,Mullen,WengRing,YPChen,Yu2015,Kim2015,Fang2015,Cava2015,Chan2016,Li2016,Bian2016,Schoop,Fang2016}.
Upon symmetry breaking, the loop can be reduced to Weyl points or a full gap~\cite{YPChen}.
Hence such a nodal-line semimetal (NLSM) may be regarded as a parent phase for Weyl semimetals and TI's.
Indeed, such transformations have been observed in concrete materials~\cite{WengRing,YPChen,Yu2015,Kim2015,Weng2015,Huang2015,Lv2015,Xu2015,Schoop}.

Characterizing and distinguishing topological phases have been a critical topic in the field.
Currently, most studies of 3D topological materials strongly rely on
angle-resolved photoemission spectroscopy (ARPES) that images the bulk and surface band structures~\cite{ex4,ex6,ex7}.
However, ARPES results in some materials may not be sufficiently clear to resolve possible avoided band crossings,
due to the narrow gaps and complex bands.
Magneto-transport could in principle offer useful information~\cite{Son2,Zhou,Burkov,HChen,SZhang,Ong,XHuang,Shekhar},
yet its indirect inference is often complicated
by alternative mechanisms, charge impurities, conventional bands etc~\cite{Goswami,YGao}.
This is in sharp contrast to the 2D case in which topological materials can
enjoy gate-tunable doping, ultra-high mobilities, and quantized Hall conductances.
Therefore, it is desirable to develop new methods for studying topological phases of 3D matter.

In this Letter, we propose new schemes for probing 3D topological materials via circular dichroism
by taking its advantages of unparalleled precision and versatile controllability~\cite{WYao,HZeng,Mak,Cao,Mak2,XXu}.
We first develop a general theory for optical transitions in 3D materials induced by circularly polarized lights,
which manifests Berry curvature effects.
Applying the theory to the NLSM, where the curvature vanishes apart from the nodal ring,
and to its derived topological phases, where the curvature spreads under symmetry breaking,
we reveal their characteristic scalings of optical absorption with photon energy.
Moreover, in optoelectronic transport, we discover a novel radial Hall effect:
charge carriers pumped by the circularly polarized light
undergo an anomalous radial flow with respect to the applied electric field,
and the azimuthal dependence leads to measurable Hall voltages at the sample surface.
Our results provide a new avenue for diagnosing topological phases of 3D matter.

\indent{\color{blue}{\em Model.}}---The essence of a NLSM can be well captured
by a minimal two-band effective model:
\begin{equation}\label{M1}
\mathcal{H}_0=\frac{1}{2m_\rho}(k_\rho^2-k_0^2)\sigma_z + v_z k_z\sigma_y\,,
\end{equation}
where $k_\rho^2=k_x^2+k_y^2$, $v_z$ is the vertical Fermi velocity,
and $m_\rho$ is the effective mass in the $\hat\rho$-direction.
Evidently, the Fermi ring is determined by $k_\rho=k_0$ in the $k_z=0$ plane.
Here a rotational symmetry around the $\hat z$-axis is assumed since it is not uncommon in crystals,
and our results below can be readily generalized to the case without such a symmetry, as commented later.
We can derive a linearized effective model around the ring as follows:
\begin{equation}\label{M2}
\mathcal{H}_0=v_\rho q_\rho \sigma_z + v_z q_z \sigma_y\,,
\end{equation}
where $(q_\rho, q_z)$ are the deviation of $(k_\rho,k_z)$ from the Fermi ring,
and $v_\rho= k_0/m_\rho$ is the radial Fermi velocity.
The two models~(\ref{M1}) and~(\ref{M2}) share the same density of states
$\nu(\epsilon)={m_\rho \epsilon}/({2\pi v_z}$), independent of the location of Fermi ring.
The linear $\epsilon$-dependence of $\nu(\epsilon)$ and the semimetal nature are reminiscent of graphene.
Indeed, the Fermi ring can be viewed as a circle of Dirac points with vanishing couplings along the azimuth.

The Fermi ring of a NLSM is often protected in at least three scenarios.
(i) In the presence of a chiral ($\Pi=\sigma_x$) symmetry,
$\Pi\mathcal{H}_0\Pi^{-1}=-\mathcal{H}_0$,
a winding number exists for any closed 1D manifold with a full energy gap~\cite{Chiu2016,FZhang}.
As an application, the Berry phase of the loop encircling the Fermi ring quantizes into $\pm\pi$~\cite{Yang2014,Mullen,WengRing,YPChen}.
Such a protection often weakens in the presence of spin-orbit couplings (SOC),
which may, though not necessarily, break the $\Pi$ symmetry.
(ii) Intriguingly, the presence of a composite symmetry
as the product of spatial inversion ($\mathcal{P}=\sigma_z$)
and time reversal ($\mathcal{T}=\mathcal{K}$) symmetries,
$(\mathcal{PT})\mathcal{H}_0(\mathcal{PT})^{-1}=\mathcal{H}_0$,
can also dictates the $\pm\pi$ quantization of the Berry phase defined in (i)~\cite{Mullen,WengRing,YPChen,Yu2015,Kim2015,Fang2015}.
Note that such a protection only exists in the absence of SOC,
as reflected by the fact $\mathcal{T}^2=1$.
The Fermi ring may enjoy dual protections, (i) and (ii), in some materials.
(iii) Mirror reflection ($\mathcal{M}_z=i\sigma_z$) symmetry,
$\mathcal{M}_z\mathcal{H}_0(k_z)\mathcal{M}_z^{-1}=\mathcal{H}_0(-k_z)$,
can also protect the Fermi ring~\cite{Chiu2016,Cava2015}.
As $\sigma_z=\pm$ becomes good quantum numbers labeling the bands in the $k_z=0$ and $\pi$ planes,
the Fermi ring can be viewed as a domain wall in momentum space
across which a band inversion occurs for a pair of $\pm$-bands.
Unlike the former two, this scenario is compatible with the presence of SOC.

For a perfect NLSM, the Berry curvature vanishes,
so does the circular dichroism that will be derived in Eq.~(\ref{DCP}).
This universal result follows from several symmetry arguments and is consistent with the above scenarios.
In scenario (i), the $\Pi$ symmetry dictates the conduction- and valence-band Berry curvature to satisfy
$\bm{\Omega}_{c}(\bm k)=\bm{\Omega}_{v}(\bm k)$~\cite{FZhang}.
Given the fact that the total Berry curvature must vanish, $\bm{\Omega}_{\alpha}(\bm k)=0$ ($\alpha=c,v$).
In scenario (ii), the $\mathcal{P}$ symmetry requires $\bm{\Omega}_{\alpha}(\bm k)=\bm{\Omega}_{\alpha}(-\bm k)$,
whereas the $\mathcal{T}$ symmetry requires $\bm{\Omega}_{\alpha}(\bm k)=-\bm{\Omega}_{\alpha}(-\bm k)$.
Hence $\bm{\Omega}_{\alpha}(\bm k)=0$ under the composite $\mathcal{PT}$ symmetry.
In scenario (iii), the above symmetries often emerge at low energy, and similar arguments apply,
although not directly imposed by the $\mathcal{M}_z$ symmetry.

Therefore, the Berry curvature becomes nontrivial
only if the symmetry protecting the Fermi ring is broken.
In this case, the Fermi ring may become fully gapped or lifted into pairs of Weyl points.
Without loss of generality, we consider the following mass term
that perturbs the Fermi ring and produces nontrivial curvature
\begin{equation}\label{gap}
\mathcal{H}_\delta=\Delta\cos(N\phi_{\bm k})\sigma_x\,,
\end{equation}
where $\phi_{\bm k}$ is the azimuthal orientation of $\bm k$.
Independent of the integer $N$, $\mathcal{H}_\delta$ breaks the $\Pi$ symmetry in scenario (i),
the $\mathcal{PT}$ symmetry in scenario (ii), and the $\mathcal{M}_z$ symmetry in scenario (iii).
For $N=0$, the Fermi ring becomes fully gapped;
the resulting phase may be a topological insulator
if $\mathcal{H}_\delta$ arises from SOC in scenario (i) or (ii),
or if the $\mathcal{T}$ symmetry is respected in scenario (iii).
For $N>0$, the Fermi ring lifts into $N$ pairs of Weyl points at the momenta where $\cos(N\phi_{\bm k})=0$.

\indent{\color{blue}{\em Circular dichroism.}}---For a circularly-polarized light,
we derive the following general optical matrix element:
\begin{equation}\label{MM}
|\mathcal{M}_\pm(\bm k)|^2=\frac{m_e^2(\epsilon_c-\epsilon_{v})^2}{2\hbar^2}
\left[\mathrm{Tr}'\!\left(R^{T}Q\,R\right)
\pm \bm{\Omega}_v\cdot\hat{ n}\right]\,,
\end{equation}
where $m_e$ is the electron mass, $\pm$ denotes the light helicity,
$Q_{ij}\!=\!\frac{1}{4(\epsilon_c-\epsilon_{v})}\!\left[\frac{\partial^2 (\epsilon_c-\epsilon_{v})}{\partial k_i\partial k_j}\!-\!\langle c|\frac{\partial^2 \mathcal{H}_0}{\partial k_i\partial k_j}|c\rangle\!+\!\langle v|\frac{\partial^2 \mathcal{H}_0}{\partial k_i\partial k_j}| v\rangle\right]$,
$\epsilon_{c,v}$ and $|c,v\rangle$ are the conduction- and valence-band energies and states,
$R$ is the rotation matrix from the light frame to the NLSM frame ($\bm k=R\bm k'$),
and $\hat n$ is the light propagation direction.
In Eq.~(\ref{MM}), all quantities  are evaluated at the same $\bm k$,
and $\mathrm{Tr}'$ is the sum over the first two diagonal matrix elements.
Evidently, the light helicity is tied with the Berry curvature.
This becomes more clearly manifested in the circular polarization:
\begin{equation}\label{DCP}
\eta(\bm k)\equiv\frac{|\mathcal{M}_+|^2-|\mathcal{M}_-|^2}{|\mathcal{M}_+|^2
+|\mathcal{M}_-|^2}=\frac{\bm{\Omega}_v\cdot\hat{ n}
}{\mathrm{Tr}'\!\left(R^{T}Q\,R\right)}\,.
\end{equation}
As we have discussed, the symmetry protecting the NLSM dictates
$\bm{\Omega}_v=0$ and hence $\eta=0$.
For the NLSM to exhibit any nontrivial circular dichroism, the symmetry must be broken,
which produces the mass term in Eq.~(\ref{gap}) and lifts the degeneracy along the Fermi ring.

\begin{figure}[t]
\includegraphics[width=\columnwidth]{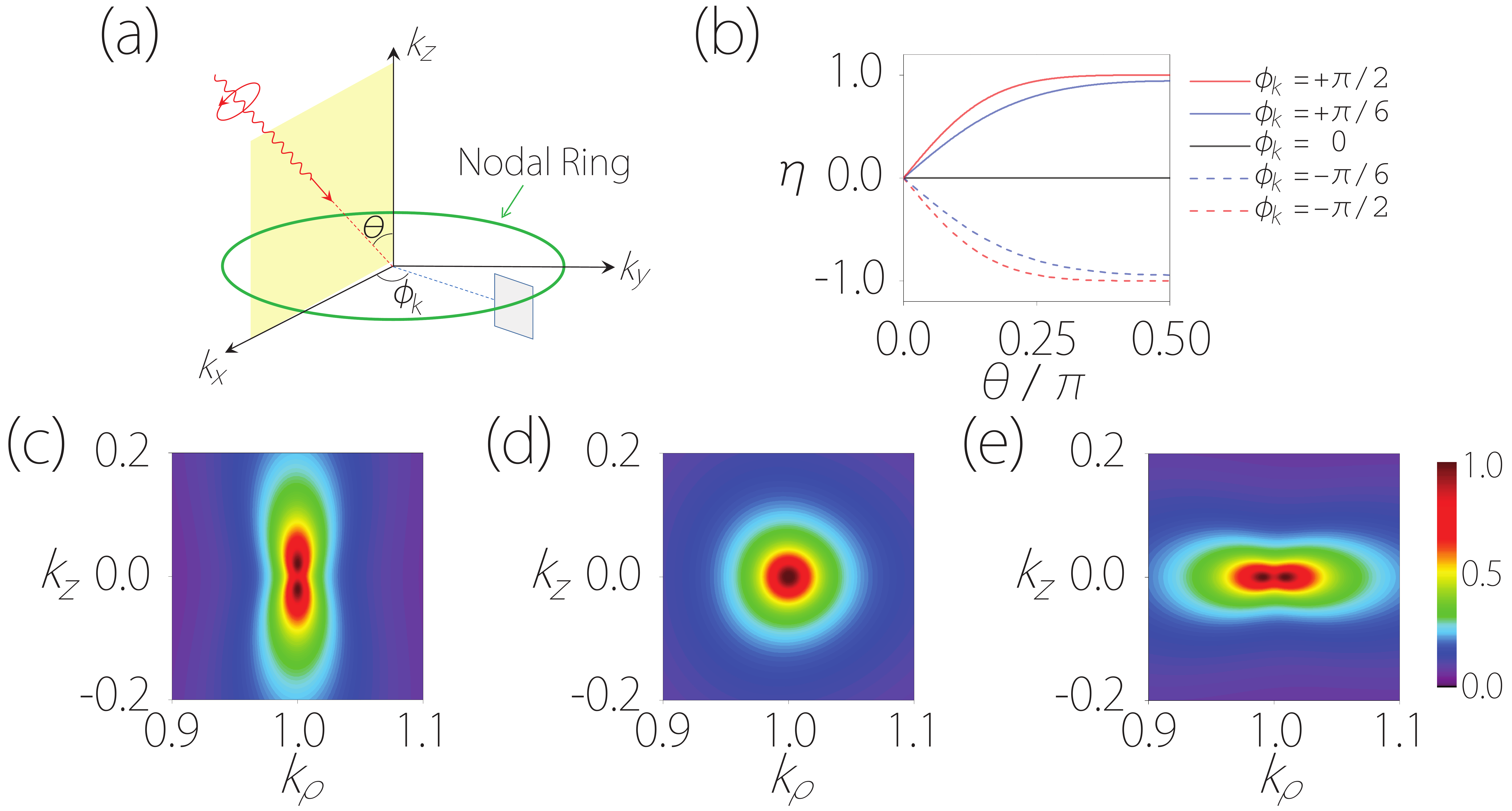}
\caption{(a) Schematics of a circularly polarized light incidence onto a NLSM.
For a fully gapped case ($N=0$), (b) circular polarization $\eta$ versus incident angle $\theta$
at different points of the original ring; (c)-(e) $\eta$ profiles for parallel incidence ($\theta=\pi/2$)
in cross sections perpendicular to the original ring with (c) $\phi_k=\pi/12$, (d) $\pi/6$, and (e) $\pi/3$.
Here $\bm k$ are measured in unit of $k_0$,
$\Delta=v_zk_0/80$, $v_\rho/v_z=1$ in (b) and $2$ in (c)-(e).}
\label{fig1}
\end{figure}

\begin{figure}[t]
\includegraphics[width=1\columnwidth]{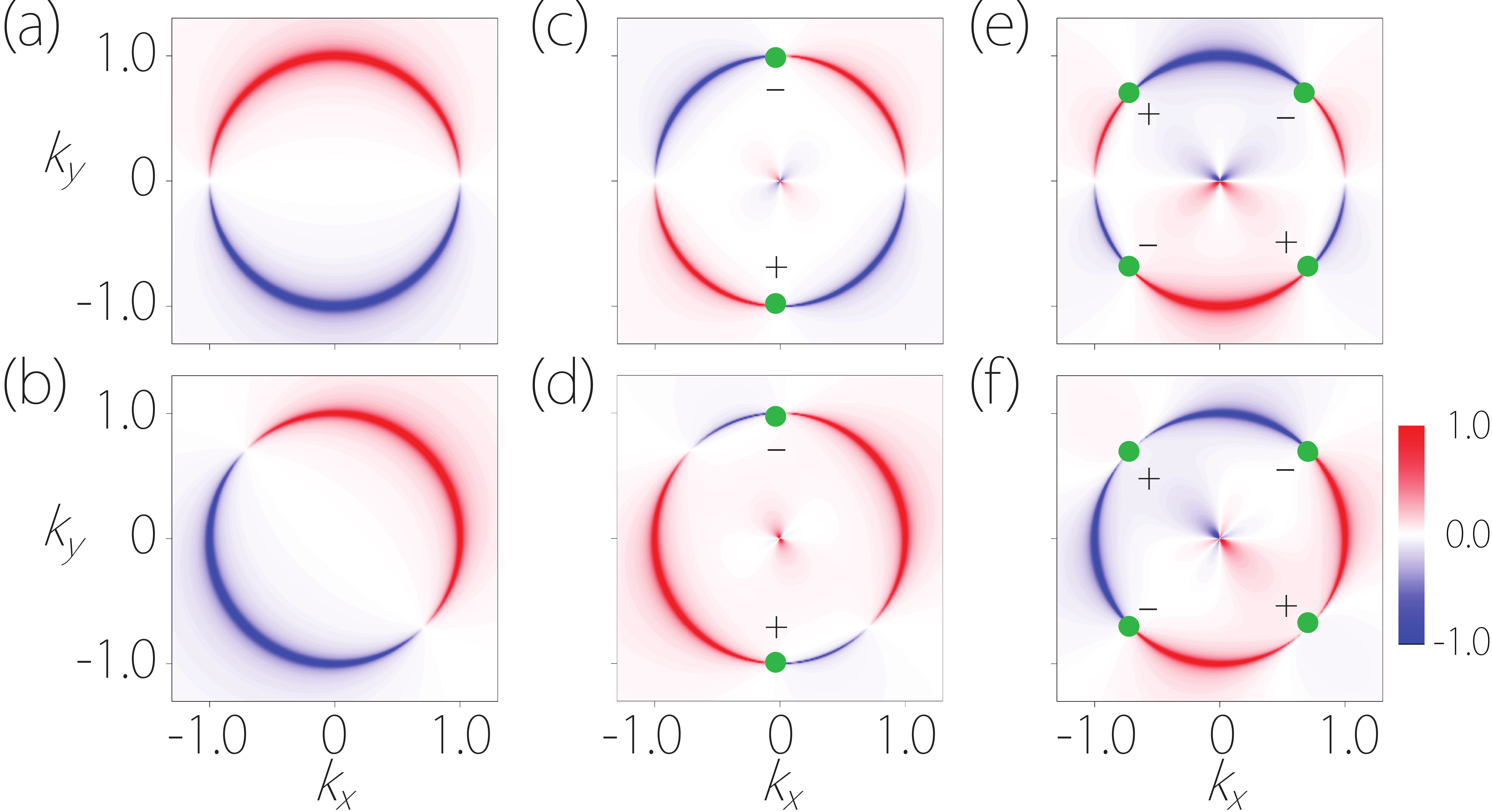}
\caption{Circular polarizations $\eta$ in the plane of the nodal ring that is gapped by Eq.~(\ref{gap})
with (a)-(b) $N=0$, (b)-(c) $N=1$, and (d)-(e) $N=2$.
{The light incident direction is $-\hat x$ in (a), (c), (e); and $\frac{-\hat x+\hat y}{\sqrt{2}}$ in (b), (d), (f).}
The green dots and $\pm$ signs mark the locations and chiralities of Weyl points.
The parameter values used here are the same as those in Fig.~\ref{fig1}.}
\label{fig2}
\end{figure}

Consider the mass term in Eq.~(\ref{gap}) with $N=0$. The Fermi ring becomes fully gapped,
and the Berry curvature peaks {\em near} the original ring in the $\hat\phi$-direction.
As $\eta\propto \bm \Omega_v\cdot\hat{ n}$ in Eq.~(\ref{DCP}),
the dichroism $\eta$ exhibits anisotropy in the direction of light incidence,
which is assumed in the $x$-$z$ plane [Fig.~\ref{fig1}(a)].
These two facts lead to that $\eta$ peaks (vanishes) for the light incidence
parallel (normal) to the plane of the original ring and strengthens (weakens)
at $|\phi_{\bm k}|=\pi/2$ ($0$) along the original ring,
as shown in Fig.~\ref{fig1}(b).
For the parallel incidence $\theta=\pi/2$, we find that
\begin{eqnarray}\label{Eq6}
\!\!\!\!\!\!\eta=\frac{ 2v_\rho v_z \Delta\epsilon_c \sin\phi_{\bm k}}
{v_\rho^2(v_z^2 k_z^2+\Delta ^2)\sin^2\phi_{\bm k}
+v_z^2[v_\rho^2 (k_\rho-k_0)^2+\Delta^2]}\,,
\end{eqnarray}
which is an odd function of $\sin\phi_{\bm k}$. As shown in  Fig.~\ref{fig2}(a),
this implies that $\eta$ in the plane of the original ring is odd
under the reflection of the $x$-$z$ incidence plane and vanishes at $\phi_{\bm k}=0$ and $\pi$.
$\eta$ in the $z$-$\rho$ plane also exhibits an interesting variation along the original ring.
Following Eq.~(\ref{Eq6}), for a 2D slice with a fixed $\phi_{\bm k}$,
when $\zeta\equiv|v_\rho\sin\phi_{\bm k}/v_z|<1$, $\eta$ has two peaks at $k_\rho=k_0$
and $k_z=\pm \frac{\Delta}{v_z\zeta}\sqrt{1-\zeta^2}$ [Fig.~\ref{fig1}(c)];
when $\zeta>1$ the two peaks are at $k_z=0$ and
$k_\rho=k_0\pm \frac{\Delta}{v_\rho}\sqrt{\zeta^2-1}$ [Fig.~\ref{fig1}(e)];
when $\zeta=1$ they merge on the original ring [Fig.~\ref{fig1}(d)].

The mass term in Eq.~(\ref{gap}) with $N=1$ gives rise to one pair of Weyl points.
Near the two Weyl points $\bm \Omega_v \propto \pm\bm q$.
In Fig.~\ref{fig2}(c), one observes that $\eta$ in the plane of the original ring
has the same sign at $\bm k$ and $-\bm k$,
a manifestation of the broken $\mathcal{T}$ symmetry
(if the original ring is around a $\mathcal{T}$-invariant momentum) or similar.
In comparison, for the $N=2$ mass term [see Fig.~\ref{fig2}(e)],
there is no overall dichroic response as the symmetry is preserved.
{For comparison, we also plot in Fig.~\ref{fig2} the cases in which the light incident direction is not $-\hat x$.}

The nontrivial $\bm k$-resolved circular polarization indicates
the selectivity of photo-induced excitations near the original ring.
Consequently, the optical absorption and anomalous transport will manifest intriguing features
that depend on the underlying mass terms in Eq.~(\ref{gap}).
These can be used to diagnose the phases and symmetries of 3D materials, as we now show.

\begin{figure}[b]
\includegraphics[width=0.9\columnwidth]{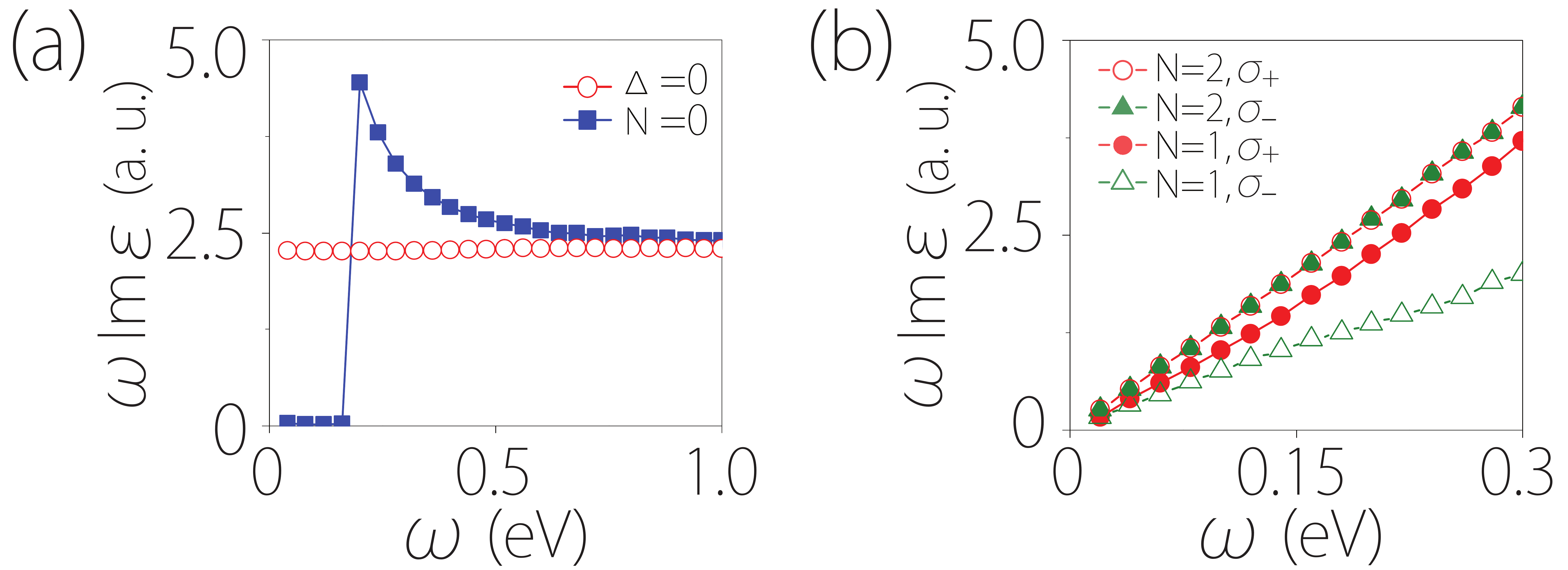}
\caption{$\omega{\rm Im}\,\varepsilon$ versus light frequency $\omega$.
(a) The case for NLSM ($\Delta=0$) and the fully gapped case ($N=0$).
(b) The two partially gapped cases with Weyl points ($N=1,2$).
We have used $k_0=0.5$~nm$^{-1}$, $k_0^2/2m_\rho = 1$~eV, $v_z/ v_\rho =1$,
$\Delta=0.1$~eV in (a), and $\Delta=0.4$~eV in (b) for illustration.}
\label{fig3}
\end{figure}

\indent{\color{blue}{\em Optical absorption.}}---The optical absorption is
related to the imaginary part of transverse dielectric function,
${\rm Im}\,\varepsilon_{\pm}(\omega)\!=\!\frac{\pi e^2}{\varepsilon_0 m^2_e\omega^2V}
\sum_{\bm k}|\mathcal{M}_{\pm}|^2\delta(\epsilon_{c}-\epsilon_{v}-\hbar\omega)$.
Thus, dielectric functions can be measured in experiment to characterize NLSM's.
For a NLSM fully gapped by Eq.~(\ref{gap}) with $N=0$,
we find that ${\rm Im}\,\varepsilon$ has a simple universal dependence on the light frequency:
\begin{equation}\label{ABS}
{\rm Im}\,\varepsilon(\omega)=\frac{\mathcal{N}}{\omega}
\left[1+\Big(\frac{2\Delta}{\hbar\omega}\Big)^2\right]\Theta(\hbar\omega-2\Delta)\,,
\end{equation}
where $\mathcal{N}$, independent of $\omega$, is a geometric factor that depends on the band parameters
and the light incident direction.
This result is valid for frequencies below which the linearized model (\ref{M2}) remains valid.
Remarkably, when $\Delta\!\ll\!\hbar\omega$, ${\rm Im}\,\varepsilon\sim 1/\omega$.
Physically, this result can be understood by noticing that a NLSM may be regarded as
a stack of 2D slices with Dirac points along the ring,
and that a 2D Dirac semimetal such as graphene has frequency-independent optical absorbance
$\propto\omega {\rm Im}\,\varepsilon$~\cite{Nair}, also referred to as $4\pi{\rm Re}\,\sigma(\omega)$.
These features are verified by our numerics in Fig.~\ref{fig3}(a).

For a NLSM partially gapped by Eq.~(\ref{gap}) with $N\neq0$,
while the high-frequency results should approach the above universal value,
the low-frequency optical transitions $\propto\omega{\rm Im}\,\varepsilon$
near the Weyl points are linear in $\omega$.
Moreover, as shown in Fig.~\ref{fig3}(b), for the case with even (odd) pairs of Weyl points,
${\rm Im}\,\varepsilon$ is independent of (sensitive to) the light helicity.
This sharp distinction can be understood in the special case
in which the center of original ring is at $k_0=0$ or $\pi$;
for the case with odd pairs of Weyl points, the $\mathcal{T}$ symmetry is broken,
and the integrated circular dichroism can survive.

\indent{\color{blue}{\em Radial Hall effect.}}---The inhomogeneous distribution of
Berry curvature in a symmetry-breaking NLSM
allows selective optical pumping of carriers in momentum space.
Such $k$-selectiveness gives rise to a radial Hall effect.
Berry curvature acts like a magnetic field in the $k$-space and, in an electric field,
induces an anomalous transverse velocity for electrons,
as reflected in the semiclassical equation of motion~\cite{Niu,XCN}:
\begin{equation}\label{EOM}
\dot{\bm r}=\frac{1}{\hbar}\frac{\partial \epsilon_c}{\partial \bm k}+\frac{e}{\hbar}{\bm E}\times \bm \Omega_{c}\,.
\end{equation}
Thus, the anomalous velocity vanishes for a NLSM and arises in the symmetry-breaking cases.

We focus on the $N=0$ case, in which at low energy
$\bm \Omega_{c}(\bm k)=-\frac{v_\rho v_z\Delta}{2\epsilon_c^3}\hat{\phi}_{\bm k}$.
An applied electric field $\bm E=E\hat{z}$ produces
anomalous velocities in the radial direction:
$\bm v_e^{\rm av}=\frac{eE\Omega_{c}}{\hbar}{\hat\rho}$ for the excited electrons
and $\bm v_h^{\rm av}=-\bm v_e^{\rm av}$ for the excited holes.
As the transverse currents from states at $k$ and $-k$ cancel each other,
the net charge Hall current vanishes.
Yet, an incident light with a circular polarization can break the symmetry
and pump nontrivial charge Hall current.
Consider a $\sigma+$ light propagating along $-\hat x$ direction with a frequency close to the band gap.
The optical transitions mainly occur for $k$-states near the original ring, and
$|\mathcal{M}_+|^2=\frac{m_e^2v_\rho^2}{2\hbar^2}(\frac{v_z}{v_\rho}+ \sin\phi_{\bm k})^2$.
It follows that the light pumps more carriers from the half ring of $k_y>0$ than the other half.
Therefore, there is a net flow of electrons (holes) in $\hat y$ ($-\hat y$) direction.
This Hall current develops a voltage detectable at surface,
and its sign flips when the light helicity is reversed.

\begin{figure}
\includegraphics[width=6cm]{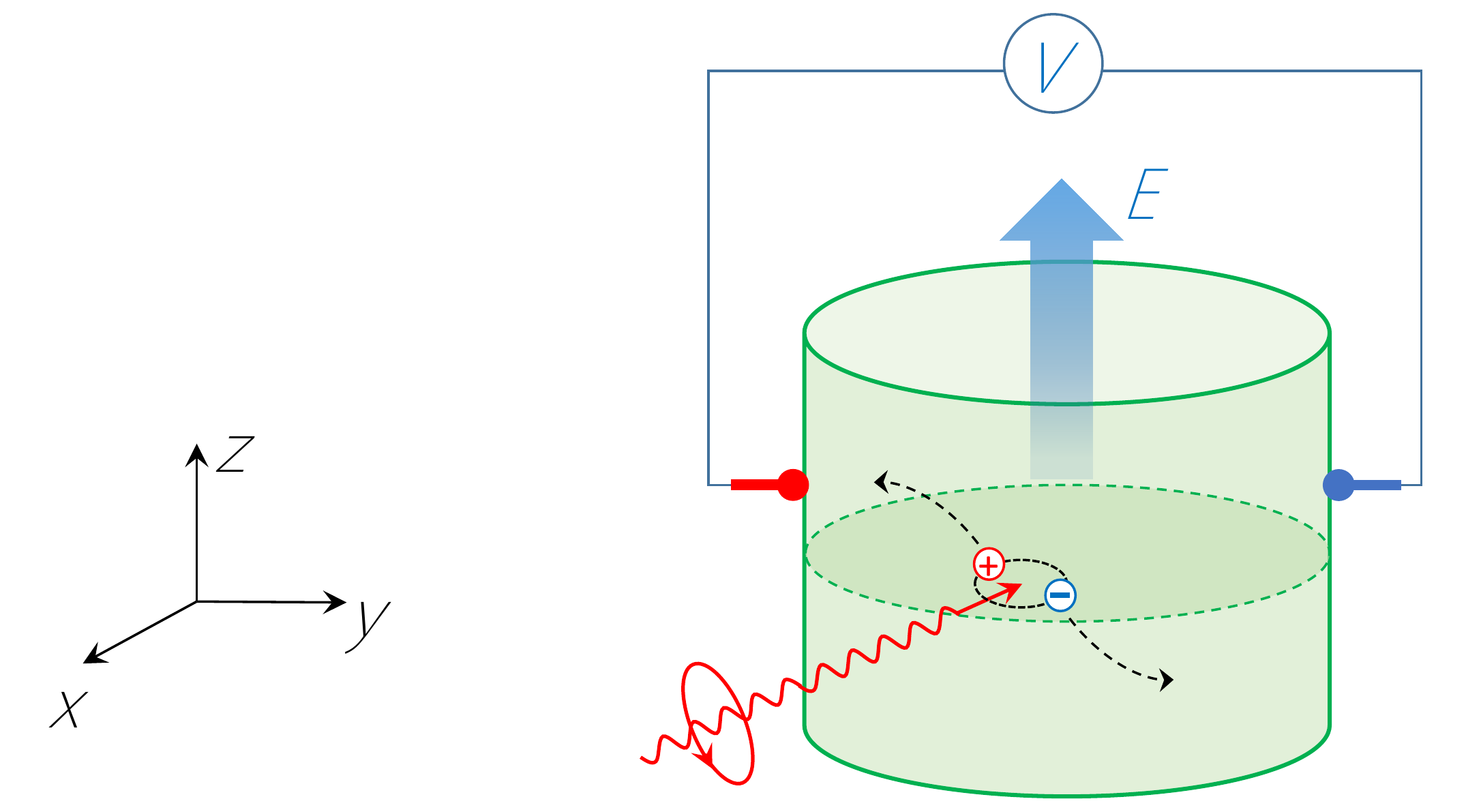}
\caption{Schematics of a radial Hall effect in a symmetry-breaking NLSM.
Electrons and holes are excited by a circularly polarized light and driven towards opposite sides of a cylindrical sample
by an electric field  perpendicular to the light incident direction.
The surface charge accumulation has strong azimuthal dependence producing Hall voltages.}
\label{fig4}
\end{figure}

Consider a cylindrical sample of radius $R$ and axis $\hat z$, as sketched in Fig.~\ref{fig4}.
The charge density accumulation $\varrho$ due to the anomalous velocities
should exhibit a transverse profile near the sample's side surface as follows:
\begin{equation}
\varrho(\rho,\phi)\propto n(\rho)\sin\phi\,,
\end{equation}
where $n(\rho)$ is the radial distribution,
and $\sin\phi$ reflects the azimuthal anisotropy.
To illustrate the factors determing $n(\rho)$,
we consider the drift-diffusion model for currents:
\begin{eqnarray}\label{current}
\bm j_\alpha=\alpha\mu_\alpha n_\alpha\bm E
+n_\alpha\bar{\bm v}_\alpha^\text{av}-D_\alpha{\bm \nabla} n_\alpha\,.
\end{eqnarray}
Here $\alpha=\pm=e,h$ label the electrons and holes, $n_\alpha$ are the carrier densities,
$\mu_\alpha$ are the mobilities, and $D_\alpha$ are the diffusion constants.
Since the carriers are mainly excited in a small region near the original ring,
the average anomalous velocities may be written as
$\bar{\bm v}_e^\text{av}=-\bar{\bm v}_h^\text{av}
=-\frac{eEv_\rho v_z\Delta}{2\hbar\bar{\epsilon}_c^3}\hat{\rho}$.
Near the sample's side surface, the carrier densities must satisfy the continuity equation:
\begin{equation}\label{CE}
\frac{\partial n_\alpha}{\partial t}=-{\bm\nabla}\cdot\bm j_\alpha-\frac{n_\alpha}{\tau_\alpha}+I\,,
\end{equation}
with the boundary condition $j_\alpha=0$ at $\rho=R$.
Here $I$ is the production rate of carriers due to light pumping,
and $\tau_\alpha$ are the carrier lifetimes taking into accounts recombination, scattering, etc.
With the ansatz that $D_\alpha$ and $\tau_\alpha$ are independent of the carrier types,
we find the steady state solution of $n(\rho)=n_e-n_h$ for $\rho\lesssim R$:
\begin{flalign}
n(\rho)\approx\left\{
\begin{array}{cc}
                2I\tau{\ell_a}/{\ell_d}\,e^{-(R-\rho)/\ell_d} &  (\ell_d\gg\ell_a) \vspace{0.05in}\\
                {I\tau}\,e^{-(R-\rho)/\ell_a} & (\ell_a\gg\ell_d) \\
\end{array} \right.
\end{flalign}
where $\ell_d\!=\!\sqrt{D\tau}$ and $\ell_a\!=\!{\bar v}^\text{av}\tau\!\propto\! E$ are length scales
characterizing diffusive transport and anomalous transport, respectively.
At small $E$ field, {\em i.e.}, in the diffusive limit, the Hall voltage of optoelectronic transport is small and linear in $E$.
At large $E$ field the anomalous transport dominates, and the signal only depends on $I\tau$ and $\phi$.

Finally, we consider the partially gapped cases.
For the mass term with $N=1$, the Fermi ring is lifted into two Weyl points along the $k_x$-axis,
which is similar to the case for the simplest Weyl semimetal without $\mathcal{T}$ symmetry.
In this case a Hall current can even be generated without light pumping.
Assuming the gap is ``inverted" inside the ring,
it follows from Eq.~(\ref{EOM}) that $\sigma_{yz}=k_0e^2/\pi h$ and $\sigma_{zx}=\sigma_{xy}=0$~\cite{HP}.
For $N=2$, the Fermi ring is lifted into four Weyl points,
which is similar to the case for the simplest Weyl semimetal with $\mathcal{T}$ symmetry.
A nonzero Hall signal would requires the pumping with a circularly-polarized light.
However, due to the gapless feature, the excited carriers would relax more easily, shortening their lifetime.
Thus, the Hall signal is expected to be much weaker than the $N=0$ case.

\indent{\color{blue}{\em Discussions.}}---We have taken the two-band model~(\ref{M1}) for demonstrating the essential ideas.
Such a treatment does not apply to the cases with additional band degeneracies,
{\em e.g.}, Dirac semimetals and centrosymmetric TI's.
In fact, for any 3D topological material with both time-reversal and inversion symmetries or with chiral symmetry,
the total valence-band Berry curvature vanishes, as aforementioned.
However,  if any of these symmetry is broken,
or if an extra symmetry offers a conserved spin or pseudospin near the original nodal ring,
their models can be reduced into several copies of our model,
and our results can be generalized straightforwardly.
In our studies, the photon energies are typically smaller than the bandwidth of model~(\ref{M1}).
This is different with the recent Floquet works~\cite{d1,d2,d3}
that tackle the high-frequency limit and ignore the optical transitions in Eq.~(\ref{MM}).

Finally, a few comments are in order.
(i) The line nodes have been assumed at Fermi energy in the above analysis.
For real materials, there may be energy variation along the nodal lines,
which can be modeled by terms like $f(k)I_{2\times2}$.
Note that our qualitative results should be unaffected,
as such terms change neither the inter-band energy difference nor the Berry curvature.
(ii) The shape of nodal line may also deviate from a perfect ring.
In the extreme case, it can be a straight line across the Brillouin zone~\cite{YPChen}.
Since our model~(\ref{M2}) applies to such a case, the same characteristics follow directly.
(iii) There may be multiple rings in the Brillouin zone.
Since the essential optical transitions only occur near the original rings,
the contribution from each ring can be treated individually as long as the rings are well separated.
(iv) Our discussion can be readily extended to elliptically-polarized light, which offers more tunability.
This is particularly useful when the unnecessary rotational symmetry in our model~(\ref{M1}) is absent.

\bibliographystyle{apsrev4-1}

\begin{thebibliography}{0}%
\makeatletter
\providecommand \@ifxundefined [1]{%
 \@ifx{#1\undefined}
}%
\providecommand \@ifnum [1]{%
 \ifnum #1\expandafter \@firstoftwo
 \else \expandafter \@secondoftwo
 \fi
}%
\providecommand \@ifx [1]{%
 \ifx #1\expandafter \@firstoftwo
 \else \expandafter \@secondoftwo
 \fi
}%
\providecommand \natexlab [1]{#1}%
\providecommand \enquote  [1]{``#1''}%
\providecommand \bibnamefont  [1]{#1}%
\providecommand \bibfnamefont [1]{#1}%
\providecommand \citenamefont [1]{#1}%
\providecommand \href@noop [0]{\@secondoftwo}%
\providecommand \href [0]{\begingroup \@sanitize@url \@href}%
\providecommand \@href[1]{\@@startlink{#1}\@@href}%
\providecommand \@@href[1]{\endgroup#1\@@endlink}%
\providecommand \@sanitize@url [0]{\catcode `\\12\catcode `\$12\catcode
  `\&12\catcode `\#12\catcode `\^12\catcode `\_12\catcode `\%12\relax}%
\providecommand \@@startlink[1]{}%
\providecommand \@@endlink[0]{}%
\providecommand \url  [0]{\begingroup\@sanitize@url \@url }%
\providecommand \@url [1]{\endgroup\@href {#1}{\urlprefix }}%
\providecommand \urlprefix  [0]{URL }%
\providecommand \Eprint [0]{\href }%
\providecommand \doibase [0]{http://dx.doi.org/}%
\providecommand \selectlanguage [0]{\@gobble}%
\providecommand \bibinfo  [0]{\@secondoftwo}%
\providecommand \bibfield  [0]{\@secondoftwo}%
\providecommand \translation [1]{[#1]}%
\providecommand \BibitemOpen [0]{}%
\providecommand \bibitemStop [0]{}%
\providecommand \bibitemNoStop [0]{.\EOS\space}%
\providecommand \EOS [0]{\spacefactor3000\relax}%
\providecommand \BibitemShut  [1]{\csname bibitem#1\endcsname}%
\let\auto@bib@innerbib\@empty
\end{thebibliography}%


\begin{thebibliography}{100}
\bibitem{Kane} M. Z. Hasan and C. L. Kane, Rev. Mod. Phys. {\bf 82}, 3045 (2010).
\bibitem{Qi}   X.-L. Qi and S.-C. Zhang, Rev. Mod. Phys. {\bf 83}, 1057 (2011).
\bibitem{Chiu2016} C.-K. Chiu, J. C. Y. Teo, A. P. Schnyder, and S. Ryu, Rev. Mod. Phys. \textbf{88}, 035005 (2016).
\bibitem{Volovik} {G. E. Volovik, \emph{The Universe in a Helium Droplet} (Clarendon Press, Oxford, 2003).}

\bibitem{Murakami} {S. Murakami, New Journal of Physics {\bf 9}, 356 (2007).}
\bibitem{Wan} {X. Wan, A. M. Turner, A. Vishwanath, and S. Y. Savrasov, Phys. Rev. B {\bf 83}, 205101 (2011).}
\bibitem{Balents}  {A. A. Burkov and L. Balents, Phys. Rev. Lett. {\bf 107}, 127205 (2011).}
\bibitem{A1} K.-Y. Yang, Y.-M. Lu, and Y. Ran, Phys. Rev. B {\bf 84}, 075129 (2011).
\bibitem{A2} G. Xu, H. Weng, Z. Wang, X. Dai, and Z. Fang, Phys. Rev. Lett. {\bf 107}, 186806 (2011).
\bibitem{Young}    S. M. Young, S. Zaheer, J. C. Y. Teo, C. L. Kane, E. J. Mele,
and A. M. Rappe, Phys. Rev. Lett. \textbf{108}, 140405 (2012).
\bibitem{A3} C. Fang, M. J. Gilbert, X. Dai, and B. A. Bernevig, Phys. Rev. Lett. {\bf 108}, 266802 (2012).
\bibitem{A4} Z. Wang, Y. Sun, X.Q. Chen, C. Franchini, G. Xu, H. Weng, X. Dai, and Z. Fang, Phys. Rev. B {\bf 85}, 195320 (2012).
\bibitem{A5} Z. Wang, H. Weng, Q. Wu, X. Dai, and Z. Fang, Phys. Rev. B {\bf 88}, 125427 (2013).
\bibitem{Zhao2013} Y. X. Zhao and Z. D. Wang, Phys. Rev. Lett. \textbf{110}, 240404 (2013).
\bibitem{BJYang}   B.-J. Yang and N. Nagaosa, Nat. Commun. \textbf{5}, 4898 (2014).
\bibitem{Yang2014} S. A. Yang, H. Pan, and F. Zhang, Phys. Rev. Lett. \textbf{113}, 046401 (2014).

\bibitem{Mullen}   K. Mullen, B. Uchoa, and D. T. Glatzhofer, Phys. Rev. Lett. {\bf 115}, 026403 (2015).
\bibitem{WengRing} H. Weng, Y. Liang, Q. Xu, R. Yu, Z. Fang, X. Dai, and Y. Kawazoe, Phys. Rev. B \textbf{92}, 045108 (2015).
\bibitem{YPChen}   Y. Chen, Y. Xie, S. A. Yang, H. Pan, F. Zhang, M. L. Cohen, and S. Zhang, Nano Lett. \textbf{15}, 6974 (2015).
\bibitem{Yu2015}   R. Yu, H. Weng, Z. Fang, X. Dai, and X. Hu, Phys. Rev. Lett.\textbf{115}, 036807 (2015).
\bibitem{Kim2015}  Y. Kim, B. J. Wieder, C. L. Kane, and A. M. Rappe, Phys. Rev. Lett. \textbf{115}, 036806 (2015).
\bibitem{Fang2015} C. Fang, Y. Chen, H. Y. Kee, and L. Fu, Phys. Rev. B \textbf{92}, 081201 (2015).
\bibitem{Cava2015} L. S. Xie, L. M. Schoop, E. M. Seibel, Q. D. Gibson, W. Xie, and R. J. Cava, APL Mater. \textbf{3}, 083602 (2015).
\bibitem{Chan2016} Y. Chan, C. Chiu, M. Chou, and A. P. Schnyder, Phys. Rev. B \textbf{93}, 205132 (2016).
\bibitem{Li2016}   R. H. Li, H. Ma, X. Cheng, S. Wang, D. Li, Z. Zhang, Y. Li, and X.-Q. Chen, Phys. Rev. Lett. \textbf{117}, 096401 (2016).
\bibitem{Bian2016} G. Bian, T. R. Chang, R. Sankar, S. Y. Xu, H. Zheng, T. Neupert, C. K. Chiu, S. M. Huang, G. Chang, I. Belopolski, D. S. Sanchez, M. Neupane, N. Alidoust, C. Liu, B. Wang, C. C. Lee, H. T. Jeng, C. Zhang, Z. Yuan, S. Jia, A. Bansil, F. Chou, H. Lin, and M. Z. Hasan, Nat. Commun. \textbf{7}, 10556 (2016).
\bibitem{Fang2016} C. Fang, H. M. Weng, X. Dai, and Z. Fang, Chin. Phys. B \textbf{25}, 117106 (2016).
\bibitem{Schoop}   L. M. Schoop, M. N. Ali, C. Stra{\ss}er, A. Topp, A. Varykhalov, D. Marchenko, V. Duppel, S. S. P. Parkin, B. V. Lotsch, and C. R. Ast, Nat. Commun. \textbf{7}, 11696 (2016).
\bibitem{Weng2015}  H. Weng, C. Fang, Z. Fang, B. A. Bernevig, and X. Dai, Phys. Rev. X \textbf{5}, 011029 (2015).
\bibitem{Huang2015} S. M. Huang, S. Y. Xu, I. Belopolski, C. C. Lee, G. Chang, B. Wang, N. Alidoust, G. Bian, M. Neupane, C. Zhang, S. Jia, A. Bansil, H. Lin, and M. Z. Hasan, Nat. Commun. \textbf{6}, 7373 (2015).
\bibitem{Lv2015}    B. Q. Lv,, H. M. Weng, B. B. Fu, X. P. Wang, H. Miao, J. Ma, P. Richard, X. C. Huang, L. X. Zhao, G. F. Chen, Z. Fang, X. Dai, T. Qian, and H. Ding, Phys. Rev. X \textbf{5}, 031013 (2015).
\bibitem{Xu2015}    S. Y. Xu, I. Belopolski, N. Alidoust, M. Neupane, G. Bian, C. Zhang, R. Sankar, G. Chang, Z. Yuan,C. C. Lee, S. M. Huang, H. Zheng, J. Ma, D. S. Sanchez,B. Wang, A. Bansil, F. Chou, P. P. Shibayev, H. Lin, S. Jia, and M. Z. Hasan, Science \textbf{349}, 613 (2015).
\bibitem{ex4} B. Q. Lv,	N. Xu, H. M. Weng, J. Z. Ma, P. Richard, X. C. Huang, L. X. Zhao, G. F. Chen, C. E. Matt, F. Bisti, V. N. Strocov, J. Mesot, Z. Fang, X. Dai, T. Qian, M. Shi, and H. Ding, Nat. Phys. {\bf 11}, 724 (2015).
\bibitem{ex6} L. X. Yang, Z. K. Liu, Y. Sun, H. Peng, H. F. Yang, T. Zhang,	B. Zhou, Y. Zhang, Y. F. Guo, M. Rahn, D. Prabhakaran,	 Z. Hussain,	S.-K. Mo, C. Felser, B. Yan, and Y. L. Chen, Nat. Phys. {\bf 11}, 728 (2015).
\bibitem{ex7} S. Y. Xu,	N. Alidoust, I. Belopolski,	Z. Yuan, G. Bian, T. R. Chang, H. Zheng, V. N. Strocov,	D. S. Sanchez, G. Chang,	 C. Zhang, D. Mou, Y. Wu, L. Huang,	C. C. Lee,	S. M. Huang, B. Wang, A. Bansil, H. T. Jeng, T. Neupert, A. Kaminski, H. Lin,	 S. Jia, and M. Z. Hasan, Nat. Phys. {\bf 11}, 748 (2015).
\bibitem{Son2}   D. T. Son and B. Z. Spivak, Phys. Rev. B {\bf 88}, 104412 (2013).
\bibitem{Zhou}   J. Zhou, H. Jiang, Q. Niu, and J. Shi, Chin. Phys. Lett. {\bf 30}, 027101 (2013).
\bibitem{Burkov} A. A. Burkov, Phys. Rev. Lett. {\bf 113}, 247203 (2014).
\bibitem{HChen}  H. Chen, Y. Gao, D. Xiao, A. H. MacDonald, and Q. Niu, arXiv:1511.02557.
\bibitem{SZhang} S.-B. Zhang, H.-Z. Lu, and S.-Q. Shen, New J. Phys. {\bf 18}, 053039 (2016).
\bibitem{Ong}    J. Xiong, S. K. Kushwaha, T. Liang, J. W. Krizan, M. Hirschberger, W. Wang, R. J. Cava, and N. P. Ong, Science {\bf 350}, 413 (2015).
\bibitem{XHuang} X. Huang, L. Zhao, Y. Long, P. Wang, D. Chen, Z. Yang, H. Liang, M. Xue, H. Weng, Z. Fang, X. Dai, and G. Chen, Phys. Rev. X {\bf 5}, 031023 (2015).
\bibitem{Shekhar} C. Shekhar, A. K. Nayak, Y. Sun, M. Schmidt, M. Nicklas, I. Leermakers, U. Zeitler, Y. Skourski, J. Wosnitza, Z. Liu, Y. Chen, W. Schnelle, H. Borrmann, Y. Grin, C. Felser, and B. Yan, Nat. Phys. {\bf 11}, 645 (2015).
\bibitem{Goswami} P. Goswami, J. H. Pixley, and S. Das Sarma, Phys. Rev. B {\bf 92}, 075205 (2015).
\bibitem{YGao}    Y. Gao, S. A. Yang, and Q. Niu, Phys. Rev. B {\bf 95}, 165135 (2017).

\bibitem{WYao}    W. Yao, D. Xiao, and Q. Niu, Phys. Rev. B {\bf 77}, 235406 (2008).
\bibitem{HZeng}   H. Zeng, J. Dai, W. Yao, D. Xiao, and X. Cui, Nat. Nanotech. {\bf 7}, 490 (2012).
\bibitem{Mak}     K. F. Mak, K. He, J. Shan, and T. F. Heinz, Nat. Nanotech. {\bf 7}, 494 (2012).
\bibitem{Cao}     T. Cao, G. Wang, W. Han, H. Ye, C. Zhu, J. Shi, Q. Niu, P. Tan, E. Wang, B. Liu, and J. Feng, Nat. Commun. {\bf 3}, 887 (2012).
\bibitem{Mak2}    K. F. Mak, K. L. McGill, J. Park, and P. L. McEuen, Science {\bf 344}, 1489 (2014).
\bibitem{XXu}     X. Xu, W. Yao, D. Xiao, and T. F. Heinz, Nat. Phys. {\bf 10}, 343 (2014).

\bibitem{FZhang}  F. Zhang, C. L. Kane, and E. J. Mele, Phys. Rev. Lett. {\bf 111}, 056403 (2013).

\bibitem{Nair} R. R. Nair, P. Blake, A. N. Grigorenko, K. S. Novoselov, T. J. Booth, T. Stauber, N. M. R. Peres, and A. K. Geim, Science {\bf 320}, 1308 (2008).

\bibitem{Niu} G. Sundaram and Q. Niu, Phys. Rev. B {\bf 59}, 14915 (1999).
\bibitem{XCN} Xiao, M.-C. Chang, and Q. Niu, Rev. Mod. Phys. {\bf 82}, 1959 (2010).
\bibitem{HP} S. A. Yang, H. Pan, and F. Zhang, Phys. Rev. Lett. {\bf 115}, 156603 (2015).

\bibitem{d1} Z. Yan and Z. Wang, Phys. Rev. Lett. {\bf 117}, 087402 (2016).
\bibitem{d2} A. Narayan, Phys. Rev. B {\bf 94}, 041409(R) (2016).
\bibitem{d3} C. Chan, Y. Oh, J. Han, and P. A. Lee, Phys. Rev. B {\bf 94}, 121106(R) (2016).



\end{thebibliography}

\end{document}